\documentclass{article}

\usepackage[square,sort,comma,numbers]{natbib}
\usepackage{relsize}
\usepackage[svgnames,dvipsnames]{xcolor}
\usepackage{amssymb}
\usepackage[T1]{fontenc}
\usepackage{mathtools}
\usepackage{breqn}
\newcommand{\sfrac}[2]{\frac{#1}{#2}}
\usepackage{braket}
\usepackage{physics}
\usepackage{tikz}
\usepackage{xparse}
\usepackage{floatrow}
\usepackage{varwidth}
\usepackage[inline]{enumitem}
\usepackage{bm}
\usepackage[unicode]{hyperref}
\usepackage[margin=1.25in,headheight=13.6pt]{geometry}
\usepackage{tasks}
\usepackage[utf8]{inputenc}

\usepackage{amsthm}

\usepackage[noabbrev,capitalize,nameinlink]{cleveref}
\usepackage{changepage}
\usepackage[normalem]{ulem}
\usepackage{booktabs}
\usepackage{array}
\usepackage{graphicx}
\usepackage{soul}
\usepackage[shortcuts]{extdash}
\usepackage{appendix}
\usepackage{pbox}
\usepackage{sectsty}
\usepackage{microtype}
\usepackage{flafter}

\hypersetup{%
  colorlinks,
  bookmarksnumbered,
  pdfencoding=unicode,
  allcolors=DarkSlateGrey
}

\newcolumntype{L}[1]{>{\raggedright\let\newline\\\arraybackslash\hspace{0pt}}m{#1}}
\newcolumntype{C}[1]{>{\centering\let\newline\\\arraybackslash\hspace{0pt}}m{#1}}
\newcolumntype{M}[1]{>{\centering\let\newline\\\arraybackslash\hspace{0pt}$}m{#1}<{$}}
\newcolumntype{R}[1]{>{\raggedleft\let\newline\\\arraybackslash\hspace{0pt}}m{#1}}

\usepackage{setspace}
\usepackage{eqparbox, etoolbox}
\newlist{rdescription}{description}{1}
\AtBeginEnvironment{rdescription}{%
\setlist[rdescription]{leftmargin =\dimexpr\eqboxwidth{Des}+\labelsep}}%

\SetLabelAlign{parright}{\strut\smash{\parbox[t]\labelwidth{\raggedleft#1}}}

\usetikzlibrary{automata, positioning, decorations.pathmorphing, calc, quotes} 

\tikzset{
    every state/.append style={
        execute at begin node=$,
        execute at end node=$
    },
    initial text = 
}

\usepackage{thmtools}
\usepackage{thm-restate}
\declaretheorem{theorem}
\declaretheorem[sibling=theorem]{lemma,corollary}

\theoremstyle{definition}
\newtheorem{fact}[theorem]{Fact}
\newtheorem{definition}[theorem]{Definition}

\crefname{fact}{fact}{facts}
\Crefname{fact}{Fact}{Facts}

\crefalias{constraint}{equation}
\crefname{constraint}{constraint}{constraints}
\Crefname{constraint}{Constraint}{Constraints}
\creflabelformat{constraint}{#2\textup{(#1)}#3} 
\crefalias{sentence}{equation}
\crefname{sentence}{sentence}{sentences}
\Crefname{sentence}{Sentence}{Sentences}
\creflabelformat{sentence}{#2\textup{(#1)}#3}
\crefalias{expression}{equation}
\crefname{expression}{expression}{expressions}
\Crefname{expression}{Expression}{Expressions}
\creflabelformat{expression}{#2\textup{(#1)}#3}

\NewDocumentEnvironment{delineate}{m}{\textcolor{cyan!70!black!}{> > > > Begin: #1 > > > >}}{\textcolor{red!70!black!}{< < < < End: #1 < < < <}}

\mathchardef\mhyphen="2D

\DeclarePairedDelimiter\paren\lparen\rparen

\newcommand{\oh}[1]{\ensuremath{\mathit{o}\paren*{#1}}}

\makeatletter
\newcommand*{\IfItalicsTF}{%
  \ifx\f@shape\my@test@it
    \expandafter\@firstoftwo
  \else
    \expandafter\@secondoftwo
  \fi
}
\newcommand*{\my@test@it}{it}
\makeatother

\newcommand{\contextsensitivemathrm}[1]{\IfItalicsTF{\mathit{#1}}{\mathrm{#1}}}

\newcommand\machineformat[1]{\ensuremath{\contextsensitivemathrm{#1}}}
\newcommand\langclassformat[1]{\ensuremath{\mathsf{#1}}}

\newcommand{\langformat}[1]{\ensuremath{\mathtt{#1}}}

\NewDocumentCommand{\DTIME}{ o m }{\langclassformat{DTIME\IfValueTF{#1}{\paren[#1]{#2}}{\paren*{#2}}}}

\NewDocumentCommand{\defineautomata}{ m m }{%
    \expandafter\NewDocumentCommand\csname#1fa\endcsname{}{\ensuremath{\machineformat{#1fa}}}%
    \expandafter\NewDocumentCommand\csname rt#1fa\endcsname{}{\ensuremath{\machineformat{rt\mhyphen#1fa}}}%
    \expandafter\NewDocumentCommand\csname o#1fa\endcsname{ o }{\ensuremath{\machineformat{1#1fa\IfValueT{##1}{\paren*{##1}}}}}%
    \expandafter\NewDocumentCommand\csname t#1fa\endcsname{ o }{\ensuremath{\machineformat{2#1fa\IfValueT{##1}{\paren*{##1}}}}}%
    \expandafter\NewDocumentCommand\csname#1tm\endcsname{}{\ensuremath{\machineformat{#1tm}}}%
    \expandafter\NewDocumentCommand\csname #2FA\endcsname{ d<> }{\ensuremath{\langclassformat{#2FA}\IfValueT{##1}{\paren*{##1}}}}%
    \expandafter\NewDocumentCommand\csname RT#2FA\endcsname{ d<> }{\ensuremath{\langclassformat{RT\mhyphen#2FA}\IfValueT{##1}{\paren*{##1}}}}%
    \expandafter\NewDocumentCommand\csname O#2FA\endcsname{ o d<> }{\ensuremath{\langclassformat{1#2FA}\IfValueTF{##1}{\paren*{##1\IfValueT{##2}{, ##2}}}{\IfValueT{##2}{\paren*{##2}}}}}%
    \expandafter\NewDocumentCommand\csname T#2FA\endcsname{ o d<> }{\ensuremath{\langclassformat{2#2FA}\IfValueTF{##1}{\paren*{##1\IfValueT{##2}{, ##2}}}{\IfValueT{##2}{\paren*{##2}}}}}%
}

\defineautomata{d}{D}
\defineautomata{n}{N}
\defineautomata{a}{A}
\defineautomata{p}{P}
\defineautomata{q}{Q}
\defineautomata{qc}{QC}

\newcommand{\ith}[2][th]{\ensuremath{#2}\text{#1}}

\newcommand\setneg[1]{\ensuremath{\overline{#1}}}

\newcommand{\PL}{\langformat{PAD}}

\newcommand{\acc}{\ensuremath{\mathrm{acc}}}
\newcommand{\rej}{\ensuremath{\mathrm{rej}}}

\newcommand{\qacc}{\ensuremath{q_{\acc}}}
\newcommand{\qrej}{\ensuremath{q_{\rej}}}
\newcommand{\sacc}{\ensuremath{s_{\acc}}}
\newcommand{\srej}{\ensuremath{s_{\rej}}}

\newcommand{\lend}{\ensuremath{\rhd}}
\newcommand{\rend}{\ensuremath{\lhd}}

\newcommand{\reject}{\emph{reject}}

\setlist{itemsep=0pt}

\newlist{observations}{enumerate}{1}
\setlist[observations]{
    label=\arabic{*}.,
    ref=\arabic{*}
}
\crefname{observationsi}{observation}{observations}
\Crefname{observationsi}{Observation}{Observations}

\newlist{differences}{enumerate}{1}
\setlist[differences]{
    label=\arabic{*}.,
    ref=\arabic{*}
}
\crefname{differencesi}{difference}{differences}
\Crefname{differencesi}{Difference}{Differences}

\newlist{strategies}{enumerate}{1}
\setlist[strategies]{
    label=Strategy \arabic{*}:,
    ref=\arabic{*},
    labelwidth=\widthof{Strategy 1:},
    leftmargin=\parindent+\labelwidth+\labelsep
}
\crefname{strategiesi}{strategy}{strategies}
\Crefname{strategiesi}{Strategy}{Strategies}

\newlist{turingenum}{enumerate}{1}
\setlist[turingenum]{
    noitemsep,
    labelsep=.5em,
    leftmargin=1em+\parindent,
    labelwidth=1em,
    label=(\Roman{*}),
    ref=\Roman{*}
}

\crefname{turingenumi}{Stg.}{Stgs.}
\Crefname{turingenumi}{Stage}{Stages}

\SetLabelAlign{Center}{\hss#1\hss}

\newcommand{\narrowfont}[3]{\scalebox{#1}[1.0]{#3}}
\newcommand{\turinglabelformat}[1]{\narrowfont{0.85}{-20}{\scriptsize#1}}
\newlength{\turinglabelgap}

\ExplSyntaxOn
\DeclareExpandableDocumentCommand{\IfNoValueOrEmptyTF}{mmm}
 {
  \IfNoValueTF{#1}{#2}
   {
    \tl_if_empty:nTF {#1} {#2} {#3}
   }
 }
\ExplSyntaxOff

\NewDocumentEnvironment{turing}{ O{} m m }
 {\IfNoValueOrEmptyTF{#1}{\setlength{\turinglabelgap}{0em}}{\setlength{\turinglabelgap}{0.5em}}\small\begin{enumerate}[labelsep=0pt,align=left,parsep=0pt,leftmargin=\widthof{\turinglabelformat{#1}}+\turinglabelgap, 
 listparindent=0pt] 
  \item[]\ignorespaces#3\\[0.5em]
  \begin{turingenum}[
    nosep,
    align=Center,
    labelwidth=\widthof{\turinglabelformat{#1}},
    labelsep=\turinglabelgap,
    leftmargin=0em 
  ]}
 {\unskip\end{turingenum}\end{enumerate}}

\usepackage{crossreftools}
\makeatletter
\newcommand{\optionaldesc}[3]{%
  \phantomsection
#1\protected@edef\@currentlabel{#1}\protected@edef\cref@currentlabel{%
    [#3][\arabic{#3}][\cref@result]%
    #1%
  }\label{#2}%
}
\makeatother

\NewDocumentCommand{\defineturingitem}{ m m }{%
    \expandafter\NewDocumentCommand\csname#1item\endcsname{ o o m }{\IfValueTF{##1}{\item[\turinglabelformat{\optionaldesc{##1}{\IfValueTF{##2}{##2}{stg:##1}}{turingenumi}}]\begin{adjustwidth}{#2}{0pt}\ignorespaces##3\end{adjustwidth}}{\item[]\begin{adjustwidth}{#2}{0pt}\ignorespaces##3\end{adjustwidth}}}%
}

\defineturingitem{t}{0em}
\defineturingitem{tt}{1em}
\defineturingitem{ttt}{2em}
\defineturingitem{tttt}{3em}
\defineturingitem{ttttt}{4em}

\makeatletter
\def\squiggly{\bgroup \markoverwith{\lower3.9\p@\hbox{\sixly \scalebox{1.2}[0.65]{\char58}}}\ULon}
\makeatother

\def\mysout{\leavevmode\bgroup\def\ULthickness{1pt}\ULdepth=-.4ex\ULset}

\newcommand{\stkout}[1]{\begingroup\ifmmode\text{\mysout{\ensuremath{#1}}}\else\mysout{#1}\fi\endgroup}

\newcommand{\utkanworry}[1]{\textcolor{red!45!black!90}{\ifmmode\smash[b]{\squiggly{#1}}\else\squiggly{#1}\fi}}

\renewcommand{\textvisiblespace}[1][.7em]{%
  \makebox[#1]{%
    \kern.07em
    \vrule height.5ex
    \hrulefill
    \vrule height.5ex
    \kern.07em
  }
}

\DeclareMathAlphabet{\mathsl}{\encodingdefault}{\familydefault}{m}{sl}
\SetMathAlphabet{\mathsl}{bold}{\encodingdefault}{\familydefault}{bx}{sl}

\usepackage{fancyhdr}
\author{A. C. Cem Say\\\href{mailto:cem.say@khas.edu.tr}{\texttt{cem.say@khas.edu.tr}}}
\title{Lower space bounds for polynomial-time quantum computation}
\date{\small \itshape
  Department of Computer Engineering,
  Kadir Has University,
  \.{I}stanbul,
  T\"{u}rkiye
  }

\newcommand{\ie}{i.e.}

\NewDocumentCommand{\padlang}{ e{_} m }{\ensuremath{\PL_{\IfValueT{#1}{#1}}\paren*{#2}}}

\NewDocumentCommand{\padpair}{ s m }{%
    \IfBooleanTF{#1}%
    {\ensuremath{\paren*{\padlang{#2}, \padlang{\setneg{#2}}}}}%
    {\ensuremath{\paren{\padlang{#2}, \padlang{\setneg{#2}}}}}%
}

\NewDocumentCommand{\branch}{ m o }{\makebox{\textsc{[#1]}\IfValueT{#2}{ Branch #2:}}}
\NewDocumentCommand{\branchpr}{ m o }{\branch{\,Probability: \ensuremath{#1}\,}[#2]}
\NewDocumentCommand{\branchperc}{ m o }{\branch{#1\% prob.}[#2]}

\NewDocumentCommand{\vartextvisiblespace}{ O{.7em} O{.7ex} }{%
  \makebox[#1]{%
    \kern.07em
    \vrule height#2
    \hrulefill
    \vrule height#2
    \kern.07em
  }
}

\newcommand{\estring}{\ensuremath{\lambda}}
\newcommand{\blanksymb}{\ensuremath{\text{\vartextvisiblespace}}}
\NewDocumentCommand{\Qpub}{ e{_} }{\ensuremath{Q_{\textnormal{pub}\IfValueT{#1}{,#1}}}}
\NewDocumentCommand{\Qpri}{ e{_} }{\ensuremath{Q_{\textnormal{pri}\IfValueT{#1}{,#1}}}}
\NewDocumentCommand{\Qcom}{ e{_} }{\ensuremath{Q_{\textnormal{com}\IfValueT{#1}{,#1}}}}
\NewDocumentCommand{\bpub}{ e{_} }{\ensuremath{b_{\textnormal{pub}\IfValueT{#1}{,#1}}}}
\NewDocumentCommand{\bpri}{ e{_} }{\ensuremath{b_{\textnormal{pri}\IfValueT{#1}{,#1}}}}

\makeatletter

\def\@testdef #1#2#3{%
  \def\reserved@a{#3}\expandafter \ifx \csname #1@#2\endcsname
 \reserved@a  \else
\typeout{^^Jlabel #2 changed:^^J%
\meaning\reserved@a^^J%
\expandafter\meaning\csname #1@#2\endcsname^^J}%
\@tempswatrue \fi}

\begin{document}

\maketitle

\begin{abstract}
\noindent 
We present new results on the minimal space requirements of problems solvable by polynomial-time quantum Turing machines. We  define the notion of Monte Carlo space complexity (where the machine is allowed to give an incorrect output or exceed the designated space budget for every input with small nonzero probability) for quantum computation, and prove that for any computable nondecreasing space bound $s(n)\in \omega(1)$, there is a language that can be recognized in polynomial time by an $s(n)$-space QTM but not by any $O(1)$-space machine in that regime. 
We also demonstrate  a sublogarithmic worst-case space hierarchy for polynomial-time quantum computation, proving  that  
 $\mathsf{BQTISP}(n^{O(1)},o(s_{i}(n)))\subsetneq \mathsf{BQTISP}(n^{O(1)},O(s_{i}(n)))$, where for each $i\ge0$,  $s_{i}(n)=\left(\frac{\log n}{(\log \log n)^i}\right)$. 

\vspace{10pt}

\end{abstract}

\section{Introduction}\label{sec:intro}

The dependence of the problem-solving capabilities of quantum computers on their time and memory budgets is an important topic of study. An algorithm is considered practically feasible only if its consumption of these two types of resources grows reasonably slowly as functions of the input length. For time complexity, polynomial bounds are universally considered reasonable in this context. At the current stage of development of quantum hardware, much stricter space budgets, corresponding to logarithmic or even sublogarithmic bounds, seem more realistic.

Although substantial quantum space and time-space lower bounds are known
for   streaming models, query complexity, and for tasks with
long outputs such as sorting and matrix multiplication
\cite{HM21,AD24,BKW26}, space lower bounds for language recognition by general quantum Turing machines (QTMs), which we focus on in this paper, are considerably rarer \cite{vMW12}.
Previous work \cite{AW02,R20,R21,S26a} on the landscape of problems that can be solved by polynomial-time QTMs under worst-case small space bounds led to the following results:

\begin{itemize}
\item Remscrim \cite{R21} proved an important theorem  establishing a tradeoff  between the runtime and memory usage of small-space quantum algorithms based on a well-known difficulty measure \cite{SB96} of the language they are supposed to recognize. Using this tool, he was able to show that no sublogarithmic-space QTM can recognize   $L_{pal}$, the language of palindromes on the alphabet $\{0,1\}$, within $2^{n^{1-\Omega(1)}}$ time. Note that $L_{pal}$ can be recognized by a deterministic Turing machine in polynomial time within $O(\log n)$ space. 
\item Say \cite{S26a,S26c} demonstrated the existence of another  ``space threshold'' by exhibiting a language which has no polynomial-time  quantum algorithm when the allowed space is in $o(\log \log n)$, but can be recognized deterministically in polynomial time  within $O(\log \log n)$ space.
    \item Unlike their classical counterparts, which can only recognize regular languages in polynomial time within $o(\log \log n)$ worst-case space \cite{DS90}, 
QTMs can recognize many nonregular languages like  
\begin{equation*}
    L_{eq}=\Set{0^n 1^n | n\geq 0}, 
\end{equation*}
as well as the word problems of all finitely generated virtually abelian groups, in polynomial expected time  using only $O(1)$ space \cite{R20}.  The two-way finite automaton with quantum and classical states (2QCFA), defined by Ambainis and Watrous \cite{AW02}, which can be viewed as a QTM that never uses its work tapes, is useful for modeling such constant-space quantum algorithms.
\end{itemize}

In this paper, we first 
examine whether giving polynomial-time QTMs a very small non-constant memory budget (i.e. a space bound of extremely slow growth in $\omega(1)\cap o(\log \log n))$ increases their computational power beyond that of 2QCFAs. For this purpose, we will adopt the following  metric \cite{KV87} of space complexity that is possibly looser than the one imposed by the worst-case usage definition employed in the studies cited above. 

\begin{definition}\label{def:mc}
    A (probabilistic or quantum) Turing machine $M$ is said to recognize a language $L$ in \textit{Monte Carlo space} $s(n)$ if there exists a constant $\varepsilon\in [0,\frac{1}{2})$ such that
    \begin{itemize}
        \item for all $w\in L$, Pr\{$M$ uses at most $s(|w|)$ work tape cells and accepts $w$\}$\geq 1-\varepsilon$, and
        \item for all $w\notin L$, Pr\{$M$ uses at most $s(|w|)$ work tape cells and rejects $w$\}$\geq 1-\varepsilon$.
    \end{itemize} 
We define the time-space complexity class $\mathsf{BQTIMCSP}(t(n),s(n))$ to be the collection of languages recognizable by QTMs that operate simultaneously in Monte Carlo space $s(n)$ and expected time $t(n)$.
\end{definition}

Our main result is
\begin{theorem}\label{thm:main}
    For any computable,  nondecreasing and unbounded function  $s(n)\in \omega(1)$,
    \[
\mathsf{BQTIMCSP}\left(n^{O(1)},O(1)\right)\subsetneq \mathsf{BQTIMCSP}\left(n^{O(1)},O(s(n))\right).
\]
\end{theorem}

In an analogous study \cite{FK94} on classical probabilistic computation, Freivalds and Karpinski proved a  result separating  classes of languages recognized by probabilistic Turing machines (PTMs) using constant space from those with very slowly growing Monte Carlo space bounds. Our technique in the proof of Theorem \ref{thm:main} differs from that of Freivalds and Karpinski in two important ways: Firstly, their constructions produce machines with exponential running times, whereas we focus on polynomial-time computation. Secondly, \cite{FK94} uses the inability of small-space PTMs to recognize $L_{pal}$ to establish the separation. Even if we relaxed our time requirements,  we would have  to identify a  different ``hard problem'' for the separation, since $L_{pal}$ is known \cite{AW02} to be recognizable by exponential-time 2QCFAs. Instead, we define and use a novel problem that is solvable in polynomial time within very small (but not constant) quantum Monte Carlo space by combining diverse ideas from work by Karpinski and Verbeek \cite{KV87} and Remscrim \cite{R21}.

We also consider the intermediate range of space bounds in $\Omega(\log \log n)\cap o(\log n)$ and show the existence of an infinite hierarchy of problems that are solvable in polynomial time and small space, with each level of the hierarchy corresponding to a different space threshold in the sense described above. In the parlance of standard simultaneous expected time and worst-case space complexity classes, we prove the following result, establishing the existence of infinitely many new space thresholds (in addition to the logarithmic, double logarithmic and constant bounds mentioned above) beyond which quantum computers can solve more problems in polynomial time:

\begin{theorem}\label{thm:hier}
    For each $i \geq 0$, 
\[
\mathsf{BQTISP}\left(n^{O(1)},o\left(\frac{\log n}{(\log \log n)^{i}}\right)\right)\subsetneq \mathsf{BQTISP}\left(n^{O(1)},O\left(\frac{\log n}{(\log \log n)^i}\right)\right).
\]
\end{theorem}

The rest of the paper is structured as follows: Section \ref{sec:defs} provides the necessary background. The proof of Theorem \ref{thm:main} is presented in Section \ref{sec:main}. Section \ref{sec:hier} contains the proof of Theorem \ref{thm:hier}. Section \ref{sec:conc} is a conclusion.

\section{Definitions and background}\label{sec:defs}


We assume that the reader is familiar with deterministic Turing machines (DTMs). For the study of space complexity, we use a multi-tape variant, with a  read-only input tape, and a single read\-/write work tape. Both tapes are accessed by dedicated two-way heads. The transition function  is restricted so that the input head never attempts to move off the tape area containing the string $\lend w \rend$, where $\lend$ and $\rend$ are special endmarker symbols, and  $w$ is the input string. At the beginning of the computation, the input head is situated on the left endmarker $\lend$, and the  work head is on the leftmost cell of the work tape, which is infinite in the other direction. All cells of the work tape are initially blank. We also assume that the work head never attempts to move left of its initial position. One disregards the ``memory'' required to store the input and to keep track of the internal state and the location of the input head, and  uses only the work tape to measure the space consumption, as will be explained below.

A quantum Turing machine  \cite{W03} can be viewed as a deterministic Turing machine augmented by  a finite quantum register (in addition to its classical register represented by its set of classical internal states), a second, quantum work tape, each cell of which holds a qubit, and a measurement register of fixed size, which will be used to store the  outcome of the latest measurement of the quantum part. The quantum work tape is accessed by a  two-way head, which is also assumed to never move to the left of its initial position on the leftmost cell.\footnote{Every QTM algorithm can be transformed into an equivalent machine which satisfies this condition by employing a simple encoding where the logical qubits $\ket{0}$ and $\ket{1}$ are represented respectively by $\ket{00}$ and $\ket{01}$ (using two tape cells), and computation begins with writing a ``left endmarker'' combination like $\ket{10}$ onto the quantum work tape. In this setup, detecting the endmarker requires measuring only the first physical qubit of any such pair, leaving the entire quantum state unchanged, including any entanglement between logical qubits.}


\begin{definition}\label{def:qtm}
A \emph{quantum Turing machine (QTM)}  is an 11-tuple  $\paren*{Q,\Sigma,K,T,r,q_0,q_{acc},q_{rej},\tau_0,\delta_Q,\delta_C}$, where
\begin{itemize}
    \item $Q$ is the finite set of classical  states,
    \item $\Sigma$ is the finite input alphabet, not containing  $\lend$ and $\rend$, 
    \item $K$ is the finite  work alphabet, including the blank symbol \blanksymb,
    \item $T$ is the finite set of possible values of the measurement register,
    \item $r$, a positive integer, is the number of qubits in the quantum register,
    \item $q_0 \in Q$ is the initial   state,
    \item $q_{acc}, q_{rej}\in Q$, such that $q_{acc} \neq q_{rej}$, are the accept and reject states, respectively,
    \item $\tau_0 \in T$ is the initial value of the measurement register, 
    \item $\delta_Q: \paren*{Q \setminus \Set{q_{acc},q_{rej}}} \times \paren*{\Sigma \cup \Set{\lend, \rend}} \times K \to \Delta$, where $\Delta$ is the set of all selective quantum operations \cite{W03} that have operator-sum representations consisting of matrices of algebraic numbers, is the quantum transition function, and 
    \item $\delta_C: \paren*{Q \setminus \Set{q_{acc},q_{rej}}} \times \paren*{\Sigma \cup \Set{\lend, \rend}} \times K \times T  \to Q \times K \times \Set{-1,0,1}^3$ is the classical transition function. 
\end{itemize}
\end{definition}

At the start of  computation,  a QTM $M$ is in state $q_0$, the measurement register contains $\tau_0$, 
and all qubits of the finite quantum register and the quantum work tape are in their zero states.  Every computational step of $M$ consists of a \textit{quantum transition}, which acts on the quantum part of the machine and updates the measurement register, and  a \textit{classical transition}, which updates the classical state, classical work tape, and the three head positions, based on the presently scanned input and classical work tape symbols, and the value in the measurement register: 
\begin{itemize}
    \item If $M$ is in some non-halting state $q$, with the input head scanning the symbol $\sigma$ and the classical work tape head scanning the symbol $\kappa$, the selective quantum operation 
$\delta_Q(q,\sigma,\kappa)$
acts on the quantum register and the qubit presently scanned by the quantum work tape head.\footnote{In this regard, our QTM definition differs from Watrous' original definition \cite{W03}, where the quantum operation depends only on the classical state $q$. As explained in \cite{S26a}, the  variant  we use here is equivalent in power  to the Watrous model, and can be specialized seamlessly to the formal definition of the 2QCFA  \cite{R21}.} This action\footnote{The concrete quantum algorithm presented in Section \ref{sec:walpha} for proving an upper bound only uses unitary transformations and projective measurements, but other nonunitary quantum operations are allowed by the formalism. This general definition ensures that our lower bounds apply for algorithms employing any physically realizable quantum operation \cite{W03,R21}.} results in those $r+1$ qubits evolving, and some output $\tau$ being stored in the measurement register with an associated probability. 
    \item The ensuing classical transition  $\delta_C(q,\sigma,\kappa, \tau)=(q',\kappa',d_I,d_W,d_Q)$ updates the classical state to $q'$, changes the presently scanned classical work tape symbol to $\kappa'$, and moves the input, classical work and quantum work heads in the directions indicated by the increments $d_I$, $d_W$, and $d_Q$, respectively.
\end{itemize}
$M$ halts with acceptance (rejection) when it enters the  state $q_{acc}$ ($q_{rej}$).

For any $\varepsilon\in[0,\frac{1}{2})$, a QTM $M$ is said to \textit{recognize} a language $L$ \textit{with error bound} $\varepsilon$ if
\begin{itemize}
    \item for each input string $w\in L$, $M$ accepts $w$ with probability at least $1-\varepsilon$, and
    \item for each input string $w\notin L$, $M$ accepts $w$ with probability at most $\varepsilon$.
\end{itemize}
Such a machine is said to \textit{run within (worst-case) space} $s(n)$ if, on any input of length $n$, at most $s(n)$ cells are scanned on both the classical and quantum work tapes throughout the computation.


A computation performed by a machine that runs within $O(1)$ space can also be performed (with no increase in runtime) by a machine which does not use its work tapes at all, and encodes all possible memory contents in its classical state set and finite quantum register. This constant\-/space model is known as the \textit{two-way finite automaton with  quantum and classical states (2QCFA)} \cite{R21}. 
Note that the original definition of the 2QCFA model in \cite{AW02} involved more restricted quantum transitions, allowing only  unitary transformations and projective measurements. Stinespring's dilation theorem \cite{HZ12} guarantees that the two formulations are equivalent in computational power, as long as both are limited to employing algebraic numbers in the transition matrices.

$\mathsf{BQTISP}(t(n),s(n))$ denotes the class of languages recognizable with bounded error by QTMs running within worst-case space $s(n)$ and in expected time $t(n)$. Throughout, time and space bounds are taken to be nondecreasing.

The computational power of QTMs in general and 2QCFAs in particular depend on the set of numbers that are allowed to appear in the matrices describing their quantum transitions \cite{SY17}. The set of algebraic numbers, used for this purpose in Definition \ref{def:qtm}, is commonly regarded as the natural choice \cite{W03,R20,R21}. The concrete algorithms to be described in this paper will use only algebraic transition amplitudes. On the other hand, our lower bound results will employ QTM and 2QCFA models that are allowed to use the full set of complex numbers in this regard to show that the proven bounds are valid for even those more powerful models. In that context, $\mathsf{BQTISP}_{\mathbb{C}}(t(n),s(n))$ will denote the counterpart of $\mathsf{BQTISP}(t(n),s(n))$ defined in terms of unrestricted-amplitude QTMs. Similarly, a 2QCFA$_{\mathbb{C}}$ is simply a 2QCFA that may employ transcendental as well as algebraic amplitudes \cite{AW02}.

\begin{definition}\label{def:dissimilar}
\cite{DS90} For any language $L$ and number $n\geq 0$, two strings $w$ and $w'$ are said to be  \emph{$(L,n)$\-/dissimilar}, written $w \not\sim_{L,n} w'$, if $\abs{w} \leq n $, $\abs{w'} \leq n$, and there exists a \emph{distinguishing string} $v$ with $\abs{wv} \leq n$, $\abs{w'v} \leq n$, and $wv \in L \iff w'v \notin L$. Let $D_{L}(n)$ be the maximum $d$ such that there exist $d$ distinct strings that are pairwise $\not\sim_{L,n}$, with the convention that $D_{L}(n)=1$ if no two strings are $(L,n)$-dissimilar. The function $D_L$ is named the \textit{nonregularity of $L$}.
\end{definition}

 The following results, which  link the nonregularity measure $D_{L}$  to the simultaneous time-space complexity of a QTM that recognizes the associated language $L$, are due to Remscrim. \cite{R21}

\begin{theorem}\label{thm:remscrimQTMtradeoff}
Suppose \( L \in \mathsf{BQTISP}_{\mathbb{C}}(t(n),s(n)) \), and suppose further that \( s(n) = o(\log \log D_L(n)) \). Then there exists a real number $ b_0 >0$ such that
\[
t(n) \in \Omega\left(2^{-b_0 s(n)} D_L(n)^{2^{-b_0 s(n)}}\right).
\]
\end{theorem}

\begin{theorem}\label{thm:tndl}
    Suppose $L$ is recognized with error bound $\varepsilon$ in expected time $t(n)$ by a 2QCFA$_{\mathbb{C}}$ that has  $d$ elements in its classical state set and $r$ qubits in its quantum register. Then there exists a natural number $n_0$ such that 
    \[
    t(n) \ge \frac{(1-2\varepsilon)^2}{16\sqrt{2}\,{(2^r)}^4 d^2}\, D_L(n)^{\frac{1}{{(2^r)}^4 d^2}}
    \]
    for all $n\geq n_0$.
\end{theorem}




\section{Very small Monte Carlo space is useful for polynomial-time QTMs}\label{sec:main}

This section comprises the proof of Theorem \ref{thm:main}, which we restate below.

\vspace{0.6em}

\noindent\textbf{Theorem \ref{thm:main}.}    \textit{For any computable,  nondecreasing and unbounded function $s(n)\in \omega(1)$,}
    \[
\mathsf{BQTIMCSP}\left(n^{O(1)},O(1)\right)\subsetneq \mathsf{BQTIMCSP}\left(n^{O(1)},O(s(n))\right).
\]

\subsection{The witness languages}


We start by defining the languages that will witness the separation of Theorem \ref{thm:main}. Let 
 $\alpha:\mathcal{\mathbb{N} \rightarrow \mathbb{N}}$ be any unbounded and nondecreasing computable function. 
A language $W_{\alpha}$ will be defined for each such $\alpha$. The members of each $W_{\alpha}$ will be two-part strings of the form $\pi\%\rho$, where 
\begin{itemize}
    \item the prefix $\pi$ is the representation of a natural number $k$, in the format to be described in Section \ref{sec:KVrep}, and
    \item the postfix $\rho$ is an encoding of a member of the word problem of  the group ${\mathbb{Z}}^k$, as will be described in Section \ref{sec:wordp}. 
\end{itemize}

\subsubsection{A method of encoding natural numbers}\label{sec:KVrep}

We  recall  a classical model of universal computation. A \textit{three-counter machine (3CM)} \cite{M61,KV87} $M$ consists of a finite-state control and three (unary) counters, named $C_1$, $C_2$ and $C_3$. If $M$ has $T$ states, its state set is $Q=\{q_1,q_2,\dots,q_T\}$, with the convention that $q_1$ is the initial state and $q_T$ is the halting state. The behavior of $M$ is governed by its transition function $\delta$. For each non-halting state $q_j$ of $M$, $\delta(q_j)$ specifies one of the following actions on one of the counters, say, $C_i$:
\begin{itemize}
    \item \textsc{inc}($i,q_k$): Increase $C_i$ by 1 and switch to the (possibly new) state $q_k$.
    \item \textsc{dec}($i,q_k,q_l$): Decrease $C_i$ by 1. If $C_i$ has reached 0, switch to state $q_k$, Otherwise, switch to state $q_l$. 
\end{itemize}
We assume that the initial state is associated only with \textsc{inc} actions. Configurations of $M$ can then be represented as strings of the form $0^j1^*2^*3^*\$$, where $q_j\in Q$ is the current state, the length of the maximal substring of $i$'s equals the number in $C_i$ for $i\in\{1,2,3\}$, and the end of each configuration is marked by a $\$$.
On input $n$ (a natural number), 3CM $M$ starts from the initial configuration $01^n\$$ (i.e. with $C_2$ and $C_3$ initialized to 0), and halts only if it reaches the state $q_{T}$. We define the output of the 3CM to be the number in its first counter $C_1$ in such a halting configuration. Viewing inputs and outputs in the manner described, 
three-counter machines can compute all partial 
computable functions of the form $f: \mathcal{\mathbb{N} \rightarrow \mathbb{N}}$ \cite{M61,KV87}.\footnote{Minsky \cite{M61} describes how any Turing machine $M_T$ 
can be simulated by a corresponding 3CM $M_C$, constructed by replacing each move of $M_T$ by a sequence of actions of $M_C$ which imitate the effects of that move.}

For any 3CM $M$ that halts on every input, let $h_{M}(n)$ denote the string which represents the history of the computation of $M$ on input $n$, that is, the concatenation of the 
strings 
describing the entire sequence of configurations attained by $M$ during this computation from start to halting. 

Given an  arbitrary unbounded and nondecreasing computable function 
 $\alpha:\mathcal{\mathbb{N} \rightarrow \mathbb{N}}$, 
 we will show how to construct a 3CM $M_{\alpha''}$ that computes another total function $\alpha''$ determined by $\alpha$. For any natural number $k$, the string $h_{M_{\alpha''}}(k)$ will then be used as a   representation of  $k$. The following argument has been adapted from \cite{KV87}.


Given $\alpha$, define a new function $\alpha'$  as follows:
\begin{equation*}
\alpha'(n)=
\begin{cases}
    0, & \text{if } \{m \mid \alpha(m)\leq n\}=\emptyset, \\
    \text{max}\{m \mid \alpha(m)\leq n\}, & \text{otherwise.}
\end{cases}
\end{equation*}

Since $\alpha$ is unbounded, nondecreasing and computable, $\alpha'$ is also unbounded, nondecreasing, computable and total. 

For any $n$, the value of $\alpha''(n)$ is the sum of the  numbers $\alpha'(0),\alpha'(1),\alpha'(2),\dots ,\alpha'(n)$.
On input 0, the 3CM 
 $M_{\alpha''}$ simply computes and outputs $\alpha'(0)$.
 On any larger input $n$, $M_{\alpha''}$ extends the computation required for $\alpha''(n-1)$ by an additional iteration computing $\alpha'(n)$.
 
We note that 
\begin{equation}\label{eq:alphah}
    \alpha'(n)\leq \alpha''(n)<|h_{M_{\alpha''}}(n)|
\end{equation}
for all $n$, since the string $h_{M_{\alpha''}}(n)$ clearly contains a proper substring $1^{\alpha''(n)}$ representing $M_{\alpha''}$'s output in the description of its final configuration.

The following fact will be useful in the proof of Lemma \ref{lem:qtm}.

\begin{lemma}\label{lem:useful}
For all $k\ge0$, $k\le \alpha(|h_{M_{\alpha''}}(k)|)$.
\end{lemma}
\begin{proof}
Assume to the contrary that  $\alpha(|h_{M_{\alpha''}}(k)|)<k$ for some $k$. The definition of $\alpha'$ then dictates that
\[
\alpha'(k)\ge |h_{M_{\alpha''}}(k)|,
\]
which leads to a contradiction with (\ref{eq:alphah}).
\end{proof}

\subsubsection{Representing word problems}\label{sec:wordp}

For any  string $w$ and symbol $\sigma$ from the same alphabet, let $\#(w,\sigma)$ denote the number of occurrences of $\sigma$ in $w$. Let $\epsilon$ denote the empty string.
 For each $k\geq 1$, the group ${\mathbb{Z}}^k$   is the direct product of $k$ copies of the group $\mathbb{Z}$ (the integers, where the group operation is addition).\footnote{For a detailed exposition of the aspects of group theory underlying this work, including formal definitions of concepts such as word problems and growth rates, we refer the reader to \cite{R21}.} Consider the alphabet 
$\Sigma_k=\{a_1,b_1,\ldots,a_k,b_k\}$, with each $a_i$ corresponding to ``+1'' in the $i$th group, and $b_i$ corresponding to the inverse of generator $a_i$. Each string $w$ in $\Sigma_k^*$ represents a $k$-tuple of integers, where the $i$th integer equals $\#(w,a_i)-\#(w,b_i)$ for each $i$.
 We will call the language
\[
W_{{\mathbb{Z}}^k}=\left\{\, w \mid w\in \Sigma_k^*, \#(w,a_i)=\#(w,b_i),\ \forall i \,\right\}
\]
(i.e. the set of all strings representing the all-0 tuple) the \textit{word problem of} ${\mathbb{Z}}^k$. 

To represent these word problems for all $k$ using a single alphabet, we will use  the following injective codings.
\begin{definition}\label{def:kappak}
    For all $k\geq 1$, the function $\kappa_k:\Sigma_k^*\to\{0,1\}^{*}$ is defined 
 such that
\begin{itemize}
    \item $\kappa_k(\epsilon)=\epsilon$,
    \item $\kappa_k(a_i)=0^i10$, 
     \item $\kappa_k(b_i)=0^i11$, and
    \item $\kappa_k(\sigma w)=\kappa_k(\sigma)\kappa_k(w)$, where $\sigma \in \Sigma_k$ and $w\in\Sigma_k^{+}$
\end{itemize}
for all $i\in\{1,\dots k\}$.
\end{definition}
For each $k$, the \textit{encoded word problem of} ${\mathbb{Z}}^k$ is then defined to be
\[
W'_{{\mathbb{Z}}^k}=\left\{\, \kappa_k(w) \mid w\in W_{{\mathbb{Z}}^k} \,\right\}.
\]

For each $\alpha$, we now define   the language
\[
W_{\alpha}=\{\pi\%\rho \mid \exists k\ge 1\;(\pi= h_{M_{\alpha''}}(k)  \land \rho\in W'_{{{\mathbb{Z}}^k}})\}
\]
on the common alphabet $\{0,1,2,3,\$,\%\}$, so that each string in $W_{\alpha}$ consists of a prefix encoding some number $k$, followed by a postfix that is a member of the encoded word problem of ${\mathbb{Z}}^k$. 

\subsection{A QTM recognizing $W_{\alpha}$}\label{sec:walpha}

\begin{lemma}\label{lem:qtm}
For any computable, nondecreasing and unbounded function $\alpha$,
\[
W_{\alpha}\in \mathsf{BQTIMCSP}\left(n^{O(1)},O(\log \alpha(n))\right).
\] 
\end{lemma}
\begin{proof}
For all $\varepsilon\in(0,\frac{1}{2})$, we will show the existence of a polynomial-time QTM $M_{\alpha,\varepsilon}$ such that
\begin{itemize}
        \item $M_{\alpha,\varepsilon}$ has a single qubit in its quantum register and never uses its quantum work tape,
        \item for all $w\in W_{\alpha}$, Pr\{$M_{\alpha,\varepsilon}$ scans $O(\log \alpha(|w|))$ cells of its classical work tape  and accepts $w$\}$\geq 1-\varepsilon$, and
        \item for all $w\notin W_{\alpha}$, Pr\{$M_{\alpha,\varepsilon}$ scans $O(\log \alpha(|w|))$ cells of its classical work tape  and rejects $w$\}$\geq 1-\varepsilon$.
    \end{itemize} 

We start by explaining the idea behind a heavily used ``quantum subroutine'' of $M_{\alpha,\varepsilon}$.
 Consider the word problem of ${\mathbb{Z}}$, i.e. the language
\[
W_{{\mathbb{Z}}}=\left\{\, w \mid  \#(w,a)=\#(w,b)\right\}
\]
on the alphabet $\{a,b\}$. There exists a polynomial-time 2QCFA algorithm $M_{W_{{\mathbb{Z}}}}$ \cite{R20} which recognizes $W_{{\mathbb{Z}}}$  with negative one-sided bounded error, that is, $M_{W_{{\mathbb{Z}}}}$ accepts members of $W_{{\mathbb{Z}}}$ with probability 1, and can be tuned (by selecting a suitable instance $M_{W_{{\mathbb{Z}}},\varepsilon}$ from an infinite set of such machines) so that it  rejects nonmembers with probability at least $1-\varepsilon$, for any desired small positive value of $\varepsilon$. $M_{W_{{\mathbb{Z}}}}$ uses a single qubit in its quantum register.

We use a technique \cite{S26a} by which the algorithm $M_{W_{{\mathbb{Z}}}}$ can be adapted to construct  subroutines (i.e., ``submachines'' that form part of the program of a larger QTM) that perform variants of the basic task of comparing the lengths of two easily identifiable subsequences embedded in the input string. In the description of $M_{\alpha,\varepsilon}$ to be presented below, the common name \textsc{EqNum} will be used to  refer to various such submachines, each of which can be thought of as running an instance of $M_{W_{{\mathbb{Z}}}}$ on a ``virtual tape''. Each such submachine will be tailored for comparing two subsequences composed of specified objects within a specific segment of the input tape. For instance, one subtask that will be performed by $M_{\alpha,\varepsilon}$ is the  comparison of the number of occurrences of two different substrings (like 010 and 011) in a  segment of its input tape demarcated by the symbol $\%$ on the left and by the  endmarker $\rend$ on the right. (Assume that such a segment of the form $\%\Set{0,1}^+\rend$ exists, and that the binary string between the two delimiters can be unambiguously parsed by $M_{\alpha,\varepsilon}$ as some concatenation of  010s, 011s, and possibly other binary substrings.  As we will see, a preprocessing stage guarantees that these conditions hold for $M_{\alpha,\varepsilon}$.)
To perform this subtask, \textsc{EqNum} starts with the input head of $M_{\alpha,\varepsilon}$ on the symbol $\%$. Whenever it scans  an occurrence of a substring on the first column of Table \ref{tab:placeholder} (by executing  one or more steps of $M_{\alpha,\varepsilon}$), \textsc{EqNum} ``feeds'' $M_{W_{{\mathbb{Z}}}}$  the corresponding symbol in the second column of the table, simulating the computational step of $M_{W_{{\mathbb{Z}}}}$ associated with that  virtual symbol.
 The classical transition of the simulated $M_{W_{{\mathbb{Z}}}}$ step will then determine the new direction of the input head. The head may then move for several $M_{\alpha,\varepsilon}$ steps (possibly skipping over binary substrings other than 010 and 011) until the nearest target substring in that direction is located. The execution of \textsc{EqNum} continues in this manner until the simulated $M_{W_{{\mathbb{Z}}}}$  reaches one of its accept or reject states. If the simulated execution of $M_{W_{{\mathbb{Z}}}}$ rejects, the \textsc{EqNum} submachine also rejects, terminating the execution of $M_{\alpha,\varepsilon}$. If the simulated $M_{W_{{\mathbb{Z}}}}$ accepts, \textsc{EqNum}  simply returns control to the next state of $M_{\alpha,\varepsilon}$. The maximum probability that \textsc{EqNum}  fails to catch a mismatch in the lengths of the compared subsequences can be tuned  by setting the error bound  of the underlying $M_{W_{{\mathbb{Z}}}}$.

\begin{table}[htb!]
    \centering
\begin{tabular}{cc}
\hline
Substring on the input tape of $M_{\alpha,\varepsilon}$ & Corresponding symbol on the virtual tape of $M_{W_{{\mathbb{Z}}}}$\\
\hline
$\%$ & \lend \\
010 & $a$ \\
011 & $b$ \\
\rend & \rend \\
\hline
\end{tabular}
    \caption{Comparing the number of occurrences of ``010'' with that of ``011'' between $\%$ and \rend}
    \label{tab:placeholder}
\end{table}

    A general template that can be used to obtain $M_{\alpha,\varepsilon}$ for any desired $(\alpha,\varepsilon)$ pair is presented in Figure \ref{fig:qtmmalpha}.  We commence the explanation of the algorithm with  the implementation of the submachine  \textsc{HistoryCheck} (Stage (A)).

\begin{figure}[htb!]
    \caption{The QTM $M_{\alpha,\varepsilon}$.  $w$ denotes the input string of length $n$.}
    \label{fig:qtmmalpha} 
    \begin{turing}[(R c)]{V}{\textit{  }}
        \titem{\textit{Reject} if  $w$ is not of the form $\pi\%\rho$, where $\pi \in \{0,1,2,3,\$\}^+$ and $\rho \in \{0,1\}^*$.}
        \titem[(A)]{
           Run \textsc{HistoryCheck} to determine whether $\pi\in H_{M_{\alpha''}}$.}
        \titem[(B)]{Write the number $k=|\texttt{inp}(\pi)|$ in binary on the classical work tape.}
        \titem[(C)]{\textit{Reject} if $\rho$ is not a proper $\kappa_k$ encoding.}
        \titem{
           For each integer $i$ from 1 to $k$, do the following:}        
            \ttitem[(D)]{
           Run \textsc{EqNum} to compare the numbers of occurrences of the codewords $0^i10$ and $0^i11$ in $\rho$.}
        \titem{\textit{Accept} if not rejected up to this point.}
    \end{turing}
\end{figure}

Let  $M$ be a 3CM that eventually halts on every input. Define the ``history language'' $H_M$ of $M$ as the set of all  computation histories of $M$ (in the format described in Section \ref{sec:KVrep}) for all positive inputs:
\[
H_M=\{h_{M}(n) | n>0\}
\]
We will show that 
\textsc{HistoryCheck} embodies a  constant-space polynomial-time quantum algorithm that employs a single qubit in its quantum register   to check whether the prefix $\pi$ of the input string is a member of the history language for the  3CM $M_{\alpha''}$ (tailored for the function $\alpha$ as specified in Section \ref{sec:KVrep}). If it detects a syntax error in $\pi$, \textsc{HistoryCheck} terminates the execution with rejection. Otherwise, it returns control to the main machine to proceed with Stage (B).

\textsc{HistoryCheck} first checks, using its finite-state control, the general
configuration syntax, that the first configuration is $01^k\$$
for some $k\geq1$, and that the last configuration is halting.
 It then sweeps $\pi$, examining each consecutive pair 
\begin{equation*}
(0^{i_0}1^{i_1}2^{i_2}3^{i_3}\$,0^{j_0}1^{j_1}2^{j_2}3^{j_3}\$)
    \end{equation*}
of configuration descriptions to see if it represents a valid transition of the machine $M_{\alpha''}$. 
The finite-state control is sufficient to trace the program flow and check whether $i_0$ and $j_0$ are related correctly. 
As for the counters, note that every such transition of $M_{\alpha''}$ requires exactly one counter value  to change by 1, and the others to remain the same. For each $c\in\Set{1,2,3}$, these conditions can be controlled by  \textsc{EqNum} ``calls'' tailored to compare the number of occurrences of the symbol $c$ between the rightmost 0 and the final $\$$ of the configuration description on the left with those in the analogous segment for the configuration on the right.  
The case where the transition function of $M_{\alpha''}$ dictates that the considered counter value should increase by 1 is handled by a submachine that feeds the symbols $\lend a$ (rather than just $\lend$) to $M_{W_{{\mathbb{Z}}}}$ whenever the leftmost 0 of the configuration on the left is scanned. \textsc{dec} actions of  $M_{\alpha''}$ are handled similarly, by introducing an additional virtual $b$ for the right delimiter of the configuration on the right.

All instances of \textsc{EqNum} (those called in \textsc{HistoryCheck} as well as those in the later Stage (D)) in $M_{\alpha,\varepsilon}$  are set to operate with the desired overall error bound $\varepsilon$.
The probability that Stage (A) fails to reject when $\pi\notin H_{M_{\alpha''}}$ is at most $\varepsilon$, since any such ``illegal'' computation history that survives \textsc{HistoryCheck} must contain at least one defect which would be detected by the corresponding    \textsc{EqNum} submachine with probability at least $1-\varepsilon$.

We define the partial function $\texttt{inp}:\{0,1,2,3,\$\}^* \rightarrow  \{1\}^*$ so that it maps any string $s$ of the form 
$(0^+1^*2^*3^*\$)^+$ 
to the (possibly empty) maximal substring of 1's to the left of the leftmost $\$$ symbol in $s$. Stage (B) simply walks the input head on $\texttt{inp}(\pi)$ and writes the length $k$ of this substring in binary notation on the classical work tape.
    
    
Lemma \ref{lem:useful} and the analysis above show that   the probability that Stage (B) writes a number that is greater than $\alpha(|\pi|)$ (which cannot exceed $\alpha(n)$ due to the fact that $\alpha$ is nondecreasing) on the work tape (and therefore exceeds the space budget $O(\log \alpha(n))$) is bounded by $\varepsilon$.

 In Stage (C), $M_{\alpha,\varepsilon}$ checks whether $\rho$ is of the form $\kappa_k(\sigma_1)\kappa_k(\sigma_2)\cdots\kappa_k(\sigma_m)$ where all the $\sigma_i$ are members of $\Sigma_k$. This can be performed in a single sweep of $\rho$ by repeatedly counting up to at most $k$. Stage (D) also uses such a counter to ignore all codewords other than $0^i10$ and $0^i11$ in $\rho$ at the $i$th iteration of the loop, and presents a virtual $a$ (resp. $b$) to the algorithm $M_{W_{{\mathbb{Z}}}}$ for each $0^i10$ (resp. $0^i11$) in $\rho$, as discussed above during the explanation of \textsc{EqNum}.\footnote{Note that $\rho$ is unambiguously ``tokenizable'' to codewords at this point, since Stage (C) has been passed.}
 
 No stage uses more than $O(\log k)$ bits. The overall runtime is polynomially bounded, because all the individual stages operate in expected polynomial time even in the low-probability case where they work with a large, incorrect value of $k$, which cannot exceed the input length $n$.
$M_{\alpha,\varepsilon}$ resets and reuses its single qubit in consecutive \textsc{EqNum} calls.

If the input string is in $W_{\alpha}$, $M_{\alpha,\varepsilon}$ accepts it with probability 1. The highest probability with which a string $w \notin W_{\alpha}$ can be accepted is at most $\varepsilon$, and this is realized if $w$ has a single defect that the associated  \textsc{EqNum} subroutine fails to catch.
\end{proof}


\subsection{No constant-space polynomial-time machine can recognize $W_{\alpha}$}

This section will culminate with the proof of the following lemma.
\begin{lemma}\label{lem:2qfacannot}
    For any computable, nondecreasing and unbounded function $\alpha$, 
    \[
    W_{\alpha}\notin \mathsf{BQTIMCSP}\left(n^{O(1)},O(1)\right).
\]
\end{lemma}

We first note that the worst-case and Monte Carlo space complexity classes coincide for constant  bounds:

\begin{lemma}\label{lem:O1esit}
    $\mathsf{BQTIMCSP}\left(n^{O(1)},O(1)\right)= \mathsf{BQTISP}\left(n^{O(1)},O(1)\right).$
\end{lemma}
\begin{proof}
    For any pair of complexity functions $t$ and $s$, $\mathsf{BQTISP}\left(t(n),s(n)\right)\subseteq\mathsf{BQTIMCSP}\left(t(n),s(n)\right)$, since a language recognized by some QTM that runs within worst-case space $s(n)$ already satisfies Definition \ref{def:mc} of membership in the associated Monte Carlo space complexity class. For the other direction, assume that some  polynomial-time QTM $M$ recognizes a language $L$ in Monte Carlo space $O(1)$. This implies that there exists a natural number $s_0$ and a real number $\varepsilon<\frac{1}{2}$ such that, on every input $w$, $M$ scans at most $s_0$ cells on its work tapes and terminates with the correct accept/reject response about the membership of $w$ in $L$ with probability at least $1-\varepsilon$. We build a 2QCFA $M_0$ that simulates $M$'s behavior on ``truncated'' work tapes: $M_0$ uses its classical states to imitate the leftmost $s_0$ cells of $M$'s classical work tape, and its finite quantum register to imitate the leftmost $s_0$ qubits in $M$'s quantum work tape. Whenever $M$ attempts to access the $(s_0+1)$st cell on any work tape, $M_0$ rejects. Clearly, $M_0$ recognizes the language $L$ with error bound $\varepsilon$, and its runtime cannot exceed that of $M$, letting one conclude that  $\mathsf{BQTIMCSP}\left(n^{O(1)},O(1)\right)\subseteq \mathsf{BQTISP}\left(n^{O(1)},O(1)\right)$.
\end{proof}

To prove Lemma \ref{lem:2qfacannot}, it is therefore sufficient to show that no polynomial-time 2QCFA can recognize $W_{\alpha}$ for any  $\alpha$ satisfying the stated conditions. 

Define the \textit{length} of an element $g$ of ${\mathbb{Z}}^k$ as the length of the shortest string in $\Sigma_k^*$ that represents the $k$-tuple $g$. The number of elements of ${\mathbb{Z}}^k$ with length at most $n$ is represented by the function $\beta_{ {\mathbb{Z}}^k}(n)$, which is called the \textit{growth rate} of  ${\mathbb{Z}}^k$.

For a pair of non-decreasing functions $f_1, f_2 : \mathbb{R}_{\ge 0} \to \mathbb{R}_{\ge 0}$, one writes
$f_1 \prec f_2$ if there exist positive real numbers $c_1, c_2$ such that
for any nonnegative real $r$,
$f_1(r) \le c_1 f_2(c_1 r + c_2) + c_2$; one writes $f_1 \sim f_2$
if both $f_1 \prec f_2$ and $f_2 \prec f_1$. 

\begin{fact}\label{fact:ZKrate}
\cite{L17} For all $k\ge1$,   $\beta_{ {\mathbb{Z}}^k}(n)\sim n^k$.    
\end{fact}

Lemma 26 of \cite{R21} establishes an upper bound on the growth rate of any finitely generated group $G$ in terms of the nonregularity of the word problem of $G$. A direct application of that lemma to the groups $\mathbb{Z}^k$ yields the following result, which will be useful in our proof of Lemma \ref{lem:no}.
\begin{fact}\label{fact:grrate}
    For any $k\geq 1$, $n\geq 0$,  $\beta_{\mathbb Z^k}(n)\le D_{W_{\mathbb Z^k}}(2n)$.
\end{fact}

\begin{lemma}\label{lem:no}
    Suppose $\alpha$ is a computable, nondecreasing and unbounded function. Then there exists no polynomial-time 2QCFA$_{\mathbb{C}}$ that recognizes the language $W_{\alpha}$.
\end{lemma}
\begin{proof}
Suppose  $W_{\alpha}$ is recognized with error bound $\varepsilon<\frac{1}{2}$ in expected time $O(n^c)$ by a 2QCFA$_{\mathbb{C}}$ $M_{\alpha}^*$ which has $r$ qubits in its quantum register for some $c,r\in\mathbb{N}$. Let $Q$ be the classical state set of $M_{\alpha}^*$. Assume without loss of generality that $M_{\alpha}^*$ never reenters the initial classical state $q_0$ after the start of computation. We will show how $M_{\alpha}^*$ can be used as a template to build a 2QCFA$_{\mathbb{C}}$ $M_k$ that recognizes $W'_{{\mathbb{Z}}^k}$ for any desired $k\geq 1$. Importantly, all those $M_k$ machines for different values of $k$ will be shown to have the same classical state set, the same quantum register size, and the same polynomial exponent $c$ in their expected runtime bound  (all independent of $k$) as  $M_{\alpha}^*$. We will conclude the proof by demonstrating that this contradicts a well-known fact about the growth rate of the group ${\mathbb{Z}}^k$.

For any $k\geq 1$,  $M_k$ is obtained by modifying $M_{\alpha}^*$ so that   $M_k$'s output on any input $w\in \{0,1\}^*$ will be identical to $M_{\alpha}^*$'s output on the input string $h_{M_{\alpha''}}(k)\%w$. When its input head is scanning any symbol other than the left endmarker $\lend$,  $M_k$ simply carries out the computational step dictated by the program of $M_{\alpha}^*$. 
When scanning $\lend$, $M_k$ performs a step  that ``summarizes'' the behavior of  $M_{\alpha}^*$ on the prefix $p_k=\lend h_{M_{\alpha''}}(k)\%$ of the input tape content as described below.

Following \cite{R21}, we define the  space 
\[
H_{\partial}=\mathbb{C}^{Q}\otimes\mathbb{C}^{2^r}. 
\]
We call a pair $(q,A)$, where $q \in Q$ and $A$ is a density operator on $\mathbb{C}^{2^r}$,
 a \textit{border state}, and identify it with the density operator $\ket{q}\!\bra{q}\otimes A$ on $H_{\partial}$.
A border state represents  the ``combined internal state'' of $M_{\alpha}^*$, consisting of its classical state and the state of its quantum register, whose respective dimensions are independent of $k$.

Consider either the initial computation of $M_\alpha^*$ within $p_k$,
starting at its left endmarker, or a subsequent traversal of $p_k$,
starting at the symbol $\%$ after the head crosses the ``border'' between $p_k$ and $w$ leftward.
Such a traversal can end by accepting, rejecting, or exiting $p_k$
rightward in some classical state. 
For every border state $(q,A)$ that occurs with positive probability
at the beginning of such a traversal, the total probability of these
outcomes is $1$, since $M_\alpha^*$ has finite expected runtime. We now justify
replacing each such traversal by a single transition of $M_k$.

The construction for summarizing traversals of a fixed prefix in
\cite[Section~3.2, second remark following Definition~12]{R21}
yields, for each classical state $q$, a selective quantum operation\footnote{This operation is obtained from the
\textit{non-truncated transfer operator} of \cite{R21},
which acts on the combined classical and quantum state space,
by fixing the entering classical state and treating the
outgoing classical state as the classical outcome.}
on the original quantum register that summarizes traversals of
$p_k$ beginning in state $q$.
Its classical outcomes record acceptance, rejection, or exit in a
specified classical state. This operation is defined on all register
states and reproduces the traversal on every reachable border state. 

Our definition of a 2QCFA$_{\mathbb{C}}$ permits this operation to be performed in
a single step when $M_k$ scans $\lend$. Its classical transition
then accepts, rejects, or moves right in the indicated classical
state. The initial traversal is summarized in the same way, with
$q_0$ distinguishing it from subsequent traversals. Thus $M_k$
uses the same classical state set $Q$ and the same $r$-qubit
register as $M_\alpha^*$.\footnote{Even if we use the more restricted 2QCFA$_{\mathbb{C}}$ definition \cite{AW02,R20} that only allows unitary transformations and projective measurements in the quantum transitions,  Stinespring's dilation theorem (Section \ref{sec:defs}) guarantees that the selective quantum operation can still be realized by a 2QCFA$_{\mathbb{C}}$ gadget whose time, bit and qubit costs are constants independent of $k$.}

Since $M_k$ spends only one step whenever it is simulating  multi-step traversals performed by $M_{\alpha}^*$ on $p_k$, its runtime on any particular input $w$ cannot exceed the runtime of $M_{\alpha}^*$ on the input $h_{M_{\alpha''}}(k)\%w$.  Note that each input string given to $M_k$  is $|p_k|-1$ symbols shorter than the input processed  during the corresponding run of   $M_{\alpha}^*$, which takes time in $O(n^c)$ by our initial assumption. Therefore, $M_k$'s runtime  as a function of its input length $n$ is in $O((|p_k|-1+n)^c)$ and so there must exist  numbers $c_0,n_k$ such that $M_k$ has at most $c_0 n^c$ expected runtime for inputs longer than $n_k$, where  $n_{k}$  is a constant that may depend on $k$. This proves our claim that $M_k$ also has runtime in $O(n^c)$.

We now plug the common error bound $\varepsilon$, runtime in $O(n^c)$,  qubit count $r$ and  classical state count $d$ of the $M_k$ machines in Theorem \ref{thm:tndl} and conclude that  there exists a natural number $n'_{k}$ depending on $k$ such that 
\begin{equation}
D_{W'_{{\mathbb{Z}}^k}}(n)  \le  c_1  n^{c{(2^r)}^4 d^2}\label{eq:nonregtime}    
\end{equation}
    for all $n\geq n'_{k}$, where $c_1$ is another constant that depends on $M_{\alpha}^*$. 

We will bound the nonregularity of $W_{\mathbb Z^k}$ using (\ref{eq:nonregtime}). We claim that 
\[
D_{W_{\mathbb Z^k}}(n)\le D_{W'_{\mathbb Z^k}}\bigl((k+2)n\bigr)
\]
for all $k\geq 1$ and all $n\ge0$. 
This is easy to see for all $n$ for which $D_{W_{\mathbb Z^k}}(n)=1$. Otherwise,
let $x_1,\dots,x_t\in\Sigma_k^*$ be pairwise $(W_{\mathbb Z^k},n)$-dissimilar, where
$t=D_{W_{\mathbb Z^k}}(n)$.
For any $x\in\Sigma_k^*$ of length at most $n$, define 
\[
x' 
= \kappa_k(x).
\]
Since every codeword for the symbols in $\Sigma_k$ has length at most $k+2$,
\[
|x_i'|\le (k+2)|x_i|\le (k+2)n.
\]

Fix $i\neq j$. 
Since $x_i$ and $x_j$ are $(W_{\mathbb Z^k},n)$-dissimilar, there exists
$v\in\Sigma_k^*$ such that
\[
|x_iv|\le n,\qquad |x_jv|\le n,
\]
and
\[
x_iv\in W_{\mathbb Z^k}\iff x_jv\notin W_{\mathbb Z^k}.
\]
Then
\[
|x_i'v'|=|\kappa_k(x_iv)|\le (k+2)n,
\qquad
|x_j'v'|=|\kappa_k(x_jv)|\le (k+2)n,
\]
and
\[
x_i'v'\in W'_{\mathbb Z^k}\iff x_iv\in W_{\mathbb Z^k},
\qquad
x_j'v'\in W'_{\mathbb Z^k}\iff x_jv\in W_{\mathbb Z^k}.
\]
Hence $x_i'$ and $x_j'$ are $(W'_{\mathbb Z^k},(k+2)n)$-dissimilar. Therefore
\[
D_{W_{\mathbb Z^k}}(n)\le D_{W'_{\mathbb Z^k}}\bigl((k+2)n\bigr).
\]
Combining this fact with (\ref{eq:nonregtime}) and letting $u={c{(2^r)}^4 d^2}$ yields
\begin{equation}\label{eq:dw2k}
    D_{W_{\mathbb Z^k}}(n)\le c_2 n^u
\end{equation}
for large $n$, where $c_2$ is a constant that depends on $k$.

Fact \ref{fact:grrate} and the inequality (\ref{eq:dw2k}) imply that, for any $k\geq 1$ and all sufficiently large $n$,  
\[\label{eq:contra}
\beta_{\mathbb Z^k}(n)\le c_k n^{u},
\]
where the constant $c_k$ depends on $k$, but $u$ does not. For all $k>u$, this contradicts Fact \ref{fact:ZKrate},
 which establishes that $\beta_{ {\mathbb{Z}}^k}(n)\sim n^k$, 
 showing that our initial assumption that a polynomial-time 2QCFA$_{\mathbb{C}}$  
 can recognize $W_{\alpha}$ is false.
\end{proof}

As stated above, Lemma \ref{lem:2qfacannot} follows from Lemmas \ref{lem:no} and \ref{lem:O1esit}. 

For any computable,  nondecreasing and unbounded function  $s(n)\in \omega(1)$, set $\alpha(n)= \text{max}\{2,\lfloor s(n)\rfloor\}$. 
Lemmas \ref{lem:qtm} and \ref{lem:2qfacannot} then constitute the proof of Theorem \ref{thm:main}.  

\section{A sublogarithmic-space hierarchy for polynomial-time QTMs}\label{sec:hier}

This section contains the proof of Theorem \ref{thm:hier}, which we restate below.

\vspace{0.6em}

\noindent\textbf{Theorem \ref{thm:hier}.}    \textit{For each $i\geq0$,}
\[
\mathsf{BQTISP}\left(n^{O(1)},o\left(\frac{\log n}{(\log \log n)^{i}}\right)\right)\subsetneq \mathsf{BQTISP}\left(n^{O(1)},O\left(\frac{\log n}{(\log \log n)^i}\right)\right).
\]

\vspace{0.6em}

We will  define a language family that will help demonstrate the separations between the consecutive levels of the hierarchy.

For any integer $l\geq 0$, let bin$(l)$ denote the shortest string representing $l$ in the binary notation. The language 
\[
L_S=\{\text{bin}(0)\$\text{bin}(1)\$\text{bin}(2)\$\cdots\$\text{bin}(k) \mid k>0\}
\]
is in the deterministic complexity class $\mathsf{SPACE}(\log \log n)$ \cite{SHL65}. 
It is straightforward to show that the length of the rightmost ``segment'' (i.e. the postfix $\text{bin}(k_n)$ following the rightmost $\$$) of any $n$-symbol member of $L_S$ is related to $n$ as 
\begin{equation}\label{eq:stlen}
\frac{1}{2}\log n\le |\text{bin}(k_n)| \le \log n.\footnote{Logarithms are to the base 2.}    
\end{equation}
A polynomial-time deterministic Turing machine can recognize $L_S$ by comparing consecutive segments bit by bit, and halting when it detects an inconsistency or when all comparisons are successful. Since each comparison involves  a counter whose value can be at most $|\text{bin}(k_n)|$, the space complexity is $O(\log \log n)$. In the proof of Lemma \ref{lem:dtm} to be presented below, this algorithm will be used by a DTM  to set some variables to their initial values while using no more than $O(\log \log n)$ bits on the work tape.

The language format to be described below is based on a similar construction in   \cite{S26b}.
Consider the alphabet $\Sigma=\Set{0,1,\$}\times \Set{0,1,\#}$. Strings written using $\Sigma$ are to be interpreted as containing two tracks, with the symbol $(\sigma_u,\sigma_l)$ indicating that the upper (resp. lower) track contains the symbol $\sigma_u$ (resp. $\sigma_l$) at that position. 

For every nonnegative integer $i$,  we define the language $L_i$  as the set of all strings $w\in \Sigma^*$ which satisfy the following conditions:
 \begin{itemize}
     \item The upper track of $w$ is some member $u_n$ of  $L_S$  which ends with the postfix $\$\text{bin}(k_n)$. 
     \item The lower track of $w$ is a member of the  set
\begin{equation}\label{eq:lowtrack}
  \left\{
  p \#^{\,|u_n|-j}
  \;\middle|\;
  p \in L_{{pal}} \cap \{0,1\}^j,\;
  j = 2^{\left\lceil\frac{|\mathrm{bin}(k_n)|}{|\mathrm{bin}(|\mathrm{bin}(k_n)|)|^i}\right\rceil}
\right\}.
\end{equation}
 \end{itemize}

 The proof of Theorem \ref{thm:hier} is completed by the following two lemmas.
\begin{lemma}\label{lem:dtm}
    For all $i\ge0$, $L_i \in \mathsf{BQTISP}\left(n^{O(1)},O\left(\frac{\log n}{(\log \log n)^i}\right)\right)$.
\end{lemma}
\begin{proof}
   We describe  a polynomial-time deterministic Turing machine $M_i$ (Figure \ref{fig:dtmmi}) that recognizes $L_i$.

\begin{figure}[htb!]
    \caption{The DTM $M_i$.  $w$ denotes the input string of length $n$.}
    \label{fig:dtmmi} 
    \begin{turing}[(R c)]{V}{\textit{  }}
        \titem[(A)]{If the  string on the upper track of $w$ is a member $u_n$ of $L_S$ with final segment $\mathrm{bin}(k_n)$, store the values $|\mathrm{bin}(k_n)|$ and $|\mathrm{bin}(|\mathrm{bin}(k_n)|)|$ in binary on the work tape. Otherwise, \textit{reject}.}
        \titem[(B)]{Compute and write the number $l=\left\lceil\frac{|\mathrm{bin}(k_n)|}{|\mathrm{bin}(|\mathrm{bin}(k_n)|)|^i}\right\rceil$ in binary on the work tape.}
        \titem[(C)]{
           Write a 1 followed by $l$ 0's to store the number $j=2^l$ in a designated area on the work tape.}
        \titem[(D)]{\textit{Reject} if the  string on the lower track of $w$ is not of the form $p \#^*$, where $p\in \Set{0,1}^j$.}
        \titem[(E)]{If $p$ is a palindrome, \textit{accept}. Otherwise,  \textit{reject}.}
      \end{turing}
\end{figure}

In Stage (A), $M_i$ runs the $O(\log \log n)$-space algorithm described above \cite{SHL65}  to  determine whether the upper track of its input is some member $u_n$ of $L_S$. If this check fails, $M_i$ halts by rejecting the input. Otherwise,  $M_i$  uses a work tape segment of $O(\log \log n)$ bits to write the length (Equation \ref{eq:stlen}) of the string $\mathrm{bin}(k_n)$. $M_i$  counts the digits of that representation to obtain an even shorter binary representation of the  number $|\mathrm{bin}(|\mathrm{bin}(k_n)|)|$ as well on the tape.  The arithmetic operations required for the computation in Stage (B) can also be performed in $O(\log \log n)$ space.

Stage (C) uses $\left\lceil\frac{|\mathrm{bin}(k_n)|}{(\lfloor\log |\mathrm{bin}(k_n)|\rfloor+1)^i}\right\rceil+1$, 
i.e. $O\left(\frac{\log n}{(\log \log n)^i}\right)$ bits to 
write the number $2^l$ on the tape. That same area is used as a counter to check whether the lower track of $w$ is of the required form in Stage (D). If control reaches Stage (E), only $O(\log |p|)$ (i.e., again $O\left(\frac{\log n}{(\log \log n)^i}\right)$) bits are sufficient to check whether $p$ is a palindrome, by comparing whether the $m$th and ($|p|-m+1$)st  symbols of $p$ are identical for all $m\in \Set{1,\dots,\lvert p\rvert}$.

$M_i$'s runtime is polynomially bounded, since its space usage is logarithmically bounded and it never enters an infinite loop. We conclude by noting that
\[
\mathsf{TISP}\left(n^{O(1)},O\left(\frac{\log n}{(\log \log n)^i}\right)\right)\subseteq\mathsf{BQTISP}\left(n^{O(1)},O\left(\frac{\log n}{(\log \log n)^i}\right)\right)
\]
for all $i$, since QTMs can simulate DTMs exactly.
\end{proof}

\begin{lemma}
For all $i\geq 0$,
$
L_i \notin 
\mathsf{BQTISP}_{\mathbb{C}}\left(
n^{O(1)},
o\left(\frac{\log n}{(\log\log n)^{i}}\right)
\right)$.

\end{lemma}

\begin{proof}We first bound the nonregularity of the language $L_i$. Let
\[
N_k=\left|\mathrm{bin}(0)\$\mathrm{bin}(1)\$\cdots\$\mathrm{bin}(k)\right|
\]
be the length of the $k$th shortest member of $L_S$. So the lengths at which $L_i$ can have member strings are precisely the numbers $N_k$,
for $k>0$.

Consider such a ``populated'' length $N_k$. By Equation~\eqref{eq:stlen},
\begin{equation}\label{eq:Nkk}
 \frac12 \log N_k \leq |\mathrm{bin}(k)| \leq \log N_k.
\end{equation}
Each member $s$ with length $N_k$ has a lower track that contains a  palindrome
$p\in\{0,1\}^{j_k}$, followed by $\#$ symbols, where
\[
j_k
=
2^{
\left\lceil
\frac{|\mathrm{bin}(k)|}{|\mathrm{bin}(|\mathrm{bin}(k)|)|^i}
\right\rceil
},
\]
which is an even number. Let $l_s$ denote the  prefix of $s$ with length  $j_k/2$, that is, $l_s$ has the left half of the $j_k$-bit palindrome underlying $s$ as its lower track. For every $s$, define $r_s$ as the  postfix that satisfies $l_s r_s=s$.

 Let $W_{N_k}$ be the set $\{w \mid w=l_s, s\in L_i\cap\Sigma^{N_k}\}$.

Since every distinct pair of strings $l_s$, $l_{s'}$ in $W_{N_k}$ are distinguished by the string $r_s$,  we have 
\begin{equation}\label{eq:mega}
D_{L_i}(N_k)\geq |W_{N_k}|
=
2^{j_k/2}
=
2^{
2^{
\left\lceil
\frac{|\mathrm{bin}(k)|}{|\mathrm{bin}(|\mathrm{bin}(k)|)|^i}
\right\rceil
-1
}
}.    
\end{equation}
Using Equation (\ref{eq:Nkk}) and the fact that 
\[
|\text{bin}(m)|=\lfloor\log m\rfloor+1
\]
for any $m>0$, 
we obtain
\[
\left\lceil
\frac{|\mathrm{bin}(k)|}{|\mathrm{bin}(|\mathrm{bin}(k)|)|^i}
\right\rceil 
\geq 
\frac{1}{2}\frac{\log N_k}{((\log \log N_k) +1)^i}\frac{(\log \log N_k)^i}{(\log \log N_k)^i},
\]
which justifies that, for any $d_0\in(0,\frac{1}{2})$,
\begin{equation}\label{eq:yada}
\left\lceil
\frac{|\mathrm{bin}(k)|}{|\mathrm{bin}(|\mathrm{bin}(k)|)|^i}
\right\rceil 
\geq 
d_0\frac{\log N_k}{(\log \log N_k)^i}    
\end{equation}
 for all sufficiently large $k$, since
\[
\frac{(\log \log N_k)^i}{((\log \log N_k) +1)^i} \longrightarrow 1
\qquad (k\to\infty)
\]
for fixed $i$.

Using Inequalities (\ref{eq:mega}) and (\ref{eq:yada}) together with the fact that
\[
\frac{\log  N_k}{(\log \log N_k)^i} \longrightarrow \infty
\qquad (k\to\infty),
\]
it is easy to show that 
there exists a constant $d_1\in (0,\frac{1}{2})$ such that, for all sufficiently large $k$,
\begin{equation}\label{eq:DL-populated}
D_{L_i}(N_k)\geq 2^{2^{d_1 \frac{\log N_k}{(\log\log N_k)^i}}}.
\end{equation}

We now extend this bound from populated lengths to all sufficiently large
lengths. The function $D_{L_i}(n)$ is nondecreasing in $n$, since a set of
strings that are pairwise $(L_i,m)$-dissimilar are also pairwise $(L_i,n)$-dissimilar for
every $n\geq m$.
Let $n$ be sufficiently large, and choose $k$ such that
\[
N_k\leq n < N_{k+1}.
\]
The gap between consecutive populated lengths is
\[
N_{k+1}-N_k = 1+|\mathrm{bin}(k+1)|.
\]
Since $k< N_k\leq n$, we have
\[
|\mathrm{bin}(k+1)|\in O(\log n),
\]
and therefore
\[
n-N_k\in O(\log n).
\]
In particular, for all sufficiently large $n$,
\[
N_k\geq \frac n2.
\]
It follows that
\[
\frac{\log N_k}{(\log\log N_k)^i}
\geq
c_0 \frac{\log n}{(\log\log n)^i}
\]
for some constant $c_0>0$ and all sufficiently large $n$. Combining this with the nondecreasingness of $D_{L_i}$ and Equation~\eqref{eq:DL-populated}, we get
\begin{equation}\label{eq:DL-all-lengths}
D_{L_i}(n)
\geq
D_{L_i}(N_k)
\geq
2^{2^{d_1 \frac{\log N_k}{(\log\log N_k)^i}}}
\geq
2^{2^{d_2 \frac{\log n}{(\log\log n)^{i}}}}    
\end{equation}
for some constant $d_2>0$ and all sufficiently large $n$.

Suppose, toward a contradiction, that
\[
L_i \in 
\mathsf{BQTISP}_{\mathbb{C}}\left(
n^{O(1)},
o\left(\frac{\log n}{(\log\log n)^{i}}\right)
\right).
\]
Then there exists a  bounded-error QTM recognizing
$L_i$ in expected time $t(n)\in n^{O(1)}$
and worst-case space 
\begin{equation}\label{eq:spbo}
s(n)< a_0\frac{\log n}{(\log\log n)^{i}}
\end{equation}
for any real number $a_0>0$, for all sufficiently large $n$.

By Equation~\eqref{eq:DL-all-lengths},
\[
\log\log D_{L_i}(n)\geq d_2 \frac{\log n}{(\log\log n)^{i}}
\]
for all sufficiently large $n$. Hence
\[
s(n)\in o(\log\log D_{L_i}(n)),
\]
i.e. $s(n)$ is  small enough for  Theorem \ref{thm:remscrimQTMtradeoff} to be applicable.
Therefore, for some constant $b_0>0$,
\[
t(n)\in
\Omega\left(
2^{-b_0s(n)}
D_{L_i}(n)^{2^{-b_0s(n)}}
\right).
\]
Employing Equation~\eqref{eq:DL-all-lengths}, we obtain
\[
t(n)
\in
\Omega\left(
2^{2^{F(n)}-b_0 s(n)}
\right),
\]
where we use the abbreviation
\[
F(n)=d_2\frac{\log n}{(\log\log n)^i}-b_0s(n).
\]

We will obtain a contradiction by showing that this lower bound for $t(n)$ is superpolynomial under the given space bound. For this purpose, it is sufficient to show that the exponent 
\[
2^{F(n)}-b_0 s(n)
\]
is in $\omega(\log n)$. 
Using Equation (\ref{eq:spbo}),
we obtain, for any positive $a_0$ and all sufficiently large \(n\),
\[
F(n)
>
d_2\frac{\log n}{(\log\log n)^i}
-
b_0 a_0\frac{\log n}{(\log\log n)^{i}}.
\]
Using a suitable value of $a_0$, one gets
\begin{equation*}
F(n)>
\frac{d_2}{2}\frac{\log n}{(\log\log n)^i},
\end{equation*}
so we have
\[
\frac{2^{F(n)}}{\log n}
>
\frac{
2^{\frac{d_2}{2}\frac{\log n}{(\log\log n)^i}}
}{\log n}.
\]
Taking logarithms of the right-hand side gives
\[
\frac{d_2}{2}\frac{\log n}{(\log\log n)^i}
-
\log\log n,
\]
which tends to infinity, since \(i\) is fixed. Therefore
\[
2^{F(n)}\in\omega(\log n).
\]
On the other hand,
\[
b_0s(n)
\in
o(\log n).
\]
Hence
\begin{equation*}\label{eq:2fn}
2^{F(n)}-b_0s(n)\in\omega(\log n).    
\end{equation*}

Having obtained a contradiction by showing that $t(n)$ has to be in $n^{\omega(1)}$, we conclude that 
\[
L_i \notin 
\mathsf{BQTISP}_{\mathbb{C}}\left(
n^{O(1)},
o\left(\frac{\log n}{(\log\log n)^{i}}\right)
\right).
\]
\end{proof}


\section{Concluding remarks}\label{sec:conc}

We conclude with some open questions.

\begin{itemize}
    \item Can Theorem \ref{thm:main} be strengthened from Monte Carlo space to worst-case space?
Specifically, is it true that for every computable, nondecreasing \(s(n)\in\omega(1)\),
\[
\mathsf{BQTISP}(n^{O(1)},O(1))
\subsetneq
\mathsf{BQTISP}(n^{O(1)},O(s(n)))?
\]
\item Theorem \ref{thm:hier} exhibits space thresholds at $\Theta(\log n/(\log\log n)^i)$ for every $i$.
We believe that similar separations can be obtained for more general space functions throughout
\[
\omega(\log\log n)\cap o(\log n).
\]
\item Our techniques are clearly limited to sublogarithmic space. How can one establish superlogarithmic  space thresholds for polynomial-time quantum computation?
\end{itemize}

\section*{AI use}

Lemma \ref{lem:no} is the result of a long sequence of ideas generated in conversations with ChatGPT 5.5, with the construction of the final contradiction in the proof being due to ChatGPT. The particular function template $\frac{\log n}{(\log \log n)^i}$ used in Theorem \ref{thm:hier} was also provided by ChatGPT. The paper benefited from repeated reviews of previous drafts by ChatGPT 6. The author takes full responsibility for the contents of the manuscript.

\section*{Acknowledgments}

The author thanks Utkan Gezer for the preparation of the LaTeX infrastructure used in the figures. Special thanks and love to dear friends G\"ulay Ayy{\i}ld{\i}z Yi\u{g}itcan and Kemal Yi\u{g}itcan, as well as their cats Tosun and Sali, who hosted our family for a month during a housing emergency as this research started.

\bibliographystyle{abbrvnat}

\bibliography{references} 

\end{document}